\documentclass[a4paper]{article}

\usepackage{ISCSLP2022}

\usepackage{multirow} 
\usepackage{cite} 
\usepackage{subfigure} 
\usepackage{makecell} 
\usepackage{float} 
\usepackage{enumitem} 
\usepackage{url} 
\usepackage[dvipsnames]{xcolor}

\title{CorrectSpeech: A Fully Automated System \\for Speech Correction and Accent Reduction}
\name{Daxin Tan$^{1*}$\thanks{$^*$This work is done during internship in Huawei Noah’s Ark Lab.}, Liqun Deng$^2$, Nianzu Zheng$^2$, Yu Ting Yeung$^2$, Xin Jiang$^2$, Xiao Chen$^2$, Tan Lee$^1$}

\address{$^1$Department of Electronic Engineering, The Chinese University of Hong Kong\\
$^2$Huawei Noah’s Ark Lab}
\email{daxintan@link.cuhk.edu.hk, tanlee@ee.cuhk.edu.hk, \{dengliqun.deng,zhengnianqu,yeung.yu.ting,Jiang.Xin,chen.xiao2\}@huawei.com}

\begin{document}

\maketitle

\begin{abstract}
This study propose a fully automated system for speech correction and accent reduction.
Consider the application scenario that a recorded speech audio contains certain errors, e.g., inappropriate words, mispronunciations, that need to be corrected. The proposed system, named CorrectSpeech, performs the correction in three steps: recognizing the recorded speech and converting it into time-stamped symbol sequence, aligning recognized symbol sequence with target text to determine locations and types of required edit operations, and generating the corrected speech. Experiments show that the quality and naturalness of corrected speech depend on the performance of speech recognition and alignment modules, as well as the granularity level of editing operations. The proposed system is evaluated on two corpora: a manually perturbed version of VCTK and L2-ARCTIC. The results demonstrate that our system is able to correct mispronunciation and reduce accent in speech recordings. Audio samples are available online for demonstration
\footnote{{https://daxintan-cuhk.github.io/CorrectSpeech/}}.

\end{abstract}


\vspace{-0.5em}
\section{Introduction}
Speech is a natural and important medium for human communication. Nowadays, speech communication through audio or video over the internet has become part of our daily life. When making a long recording, for example, to give a public speech, to tell a story or to record an online lecture, we tend to make mistakes or errors. Common speech errors may include mispronounced words, missing words, stuttering or repetition. A tool for speech editing and correction is valuable to correct mispronunciation and improve fluency. The tool also saves our time and effort as we do not need to redo the whole recordings. 

A few studies have explored the field of speech editing. The VOCO system \cite{jin2017voco} leveraged a framework of unit selection speech synthesis, combined with a voice conversion module. In \cite{morrison2021context}, context-aware prosody correction was applied to speech segments extracted from the same speaker. 
In \cite{tan2021editspeech}, the EditSpeech system made the first attempt to applying neural text-to-speech models for text-based speech editing.
The EditSpeech system employed the partial inference and bidirectional fusion mechanisms to achieve high naturalness of edited speech. A system named CampNet was described in \cite{wang2022campnet}, which adopted the method of context-aware mask prediction and was built upon a transformer framework. In \cite{tae2021editts}, score-based generative modeling was adopted to deal with speech editing problem. 

In most cases, users are required to explicitly specify the edit locations and edit operations in editing process. In a typical scenario, users may have the desired scripts in hand, which contain text content that they intend to speak or express. A speech editing tool with the ability to accomplish necessary editing and correction tasks in a fully automated way is desirable. 
This motivation drives us to propose CorrectSpeech.
Given a speech utterance and the intended text content, CorrectSpeech is able to recognize the text content in the original speech, align the recognized text with the target text, determine edit locations with operations and complete the editing actions automatically. CorrectSpeech is based on the design of EditSpeech, due to its simplicity and effectiveness.

Accent reduction is another active research field in speech synthesis.
Specifically, speech accent reduction aims to endow accented speech uttered by non-native speakers with normal pronunciation pattern of native speakers, while keeping the speaker identity unchanged. Most of previous works in accent reduction follow the paradigm of voice conversion \cite{zhao2019using, zhao2019foreign,zhao2021converting, liu2020end, ding2022accentron, wang2021accent} and generate accent-reduced speech in utterance level. CorrectSpeech provides a new paradigm to deal with accent reduction problem. Instead of synthesizing a completely new speech utterance, CorrectSpeech detects and modifies the heavy-accented speech segments from the original speech utterance.

This paper is organized as follows. In the next Section, we introduce the design of CorrectSpeech. In Section \ref{sec:experiment}, we present our experimental setup. In Section \ref{sec:results}, we analyse the experimental results. 
Finally, we conclude our work in Section \ref{sec:conclusion}. 

The contributions of this paper are summarized as follows:
\begin{itemize}[leftmargin=*]
\item We develop and implement the CorrectSpeech system, which is able to correct speech with mispronunciation in a fully automated way. 

\item We carry out ablation study to examine how the choices of speech recognition and alignment modules and correction methods influence the correction performance of the overall system.
\item Experiments on L2-ARCTIC demonstrate that we may apply CorrectSpeech system for accent reduction scenario.
\end{itemize}

\begin{figure}[h]
  \centering
  \includegraphics[width=0.75\linewidth, trim=75 270 180 50]{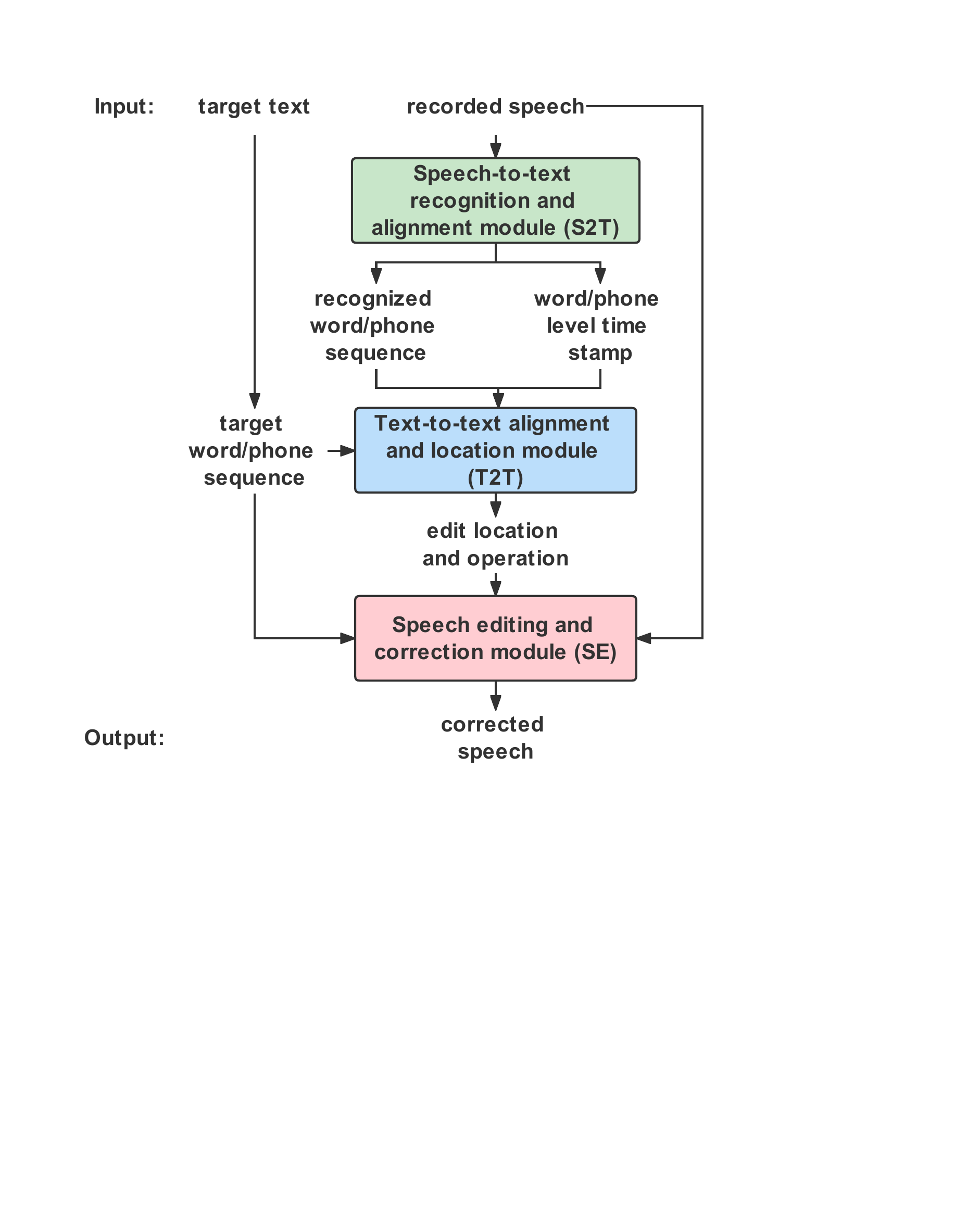}
  \caption{Overview of the CorrectSpeech system.}
  \label{fig:system_overview}
  \vspace{-1em}
\end{figure}

\vspace{-0.5em}
\section{The proposed system}

\subsection{Overview}
Figure \ref{fig:system_overview} gives an overview of the proposed CorrectSpeech. The system comprises a speech-to-text recognition and alignment module (S2T), a text-to-text alignment and location module (T2T) and a speech editing and correction module (SE).

\subsection{Speech-to-text recognition and alignment (S2T)} \label{sec:S2T}
We apply S2T module to perform recognition and alignment from speech to text. We have explored two S2T settings.

\begin{description}[leftmargin=*]
\item[F-Align] This setting consists of a byte pair encoding  (BPE) \cite{shibata1999byte} based automatic speech recognition (ASR) model for recognizing word sequence from speech. Then we apply a grapheme-to-phoneme (G2P) module to convert the word sequence to phone sequence. Finally, we apply a forced aligner to align the word sequence and the phone sequence with the speech to obtain word-level and phone-level time stamps.
\item[CTC-Align] This setting consists of a phone-based connectionist temporal classification (CTC) \cite{graves2006connectionist} ASR model for recognizing phone sequence from speech. We apply greedy decoding to obtain the recognition result. For phone-based CTC decoding, the output of each frame is either a phone label or a ``blank" label. By counting the frame index, we obtain the time-stamp of each phone, i.e. phone-level time stamps from the speech.
\end{description}

Note that the first setting F-Align follows a pipeline, while the second setting CTC-Align is an end-to-end model. In F-Align, we obtain a word sequence, a phone sequence, word-level and phone-level time stamps of the speech. In CTC-Align, we obtain a phone sequence and phone-level time stamps.

\begin{figure}[h]
  \centering
  \includegraphics[width=\linewidth, trim=40 250 35 50]{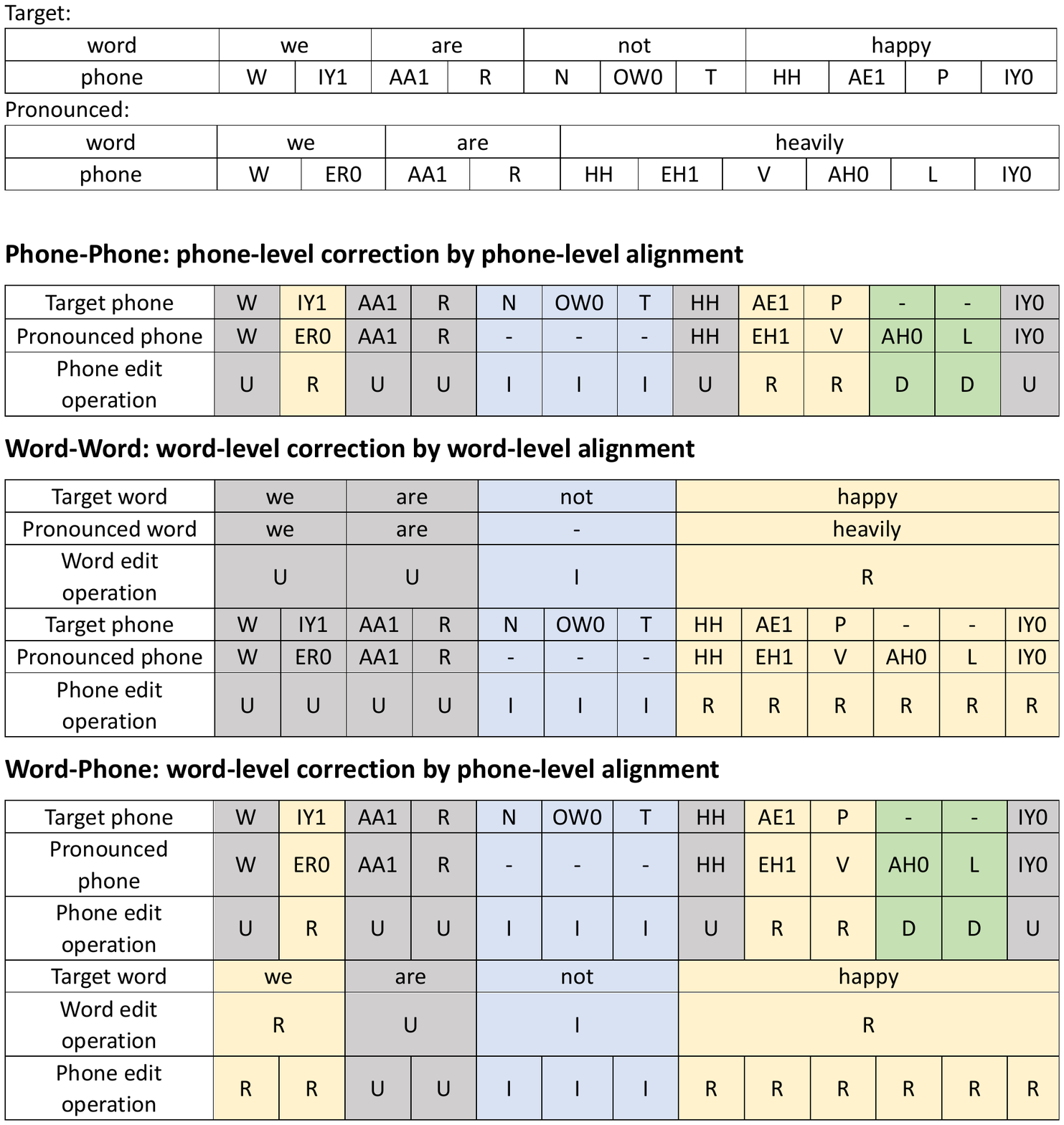}
  \caption{An example of comparison among different correction methods. ``U'', ``I'', ``R'', ``D'' represents ``unchange'', ``insert'', ``replace'' and ``delete'' respectively.}
  \label{fig:correction_method_detail}
  \vspace{-1em}
\end{figure}

\subsection{Text-to-text alignment and location (T2T)}  \label{sec:T2T}
T2T module aims to perform alignment between recognized text and target text. We consider two alignment levels: word-level and phone-level. We apply Needleman–Wunsch algorithm \cite{needleman1970general} for text alignment. The alignment process computes the minimum number of edit operations, including insertion, replacement and deletion, to convert one sequence to the other.

CorrectSpeech determines edit locations according to text-to-text alignment results. We explore three different correction methods based on different alignment levels. 

\begin{description}[leftmargin=*]
\item[Phone-Phone] Phone-level correction by phone-level alignment
\item[Word-Word] Word-level correction by word-level alignment
\item[Word-Phone] Word-level correction by phone-level alignment
\end{description}

For Phone-Phone and Word-Word, the correction granularity is the same as alignment level. Edit operation is directly derived from alignment result. We assign one of the edit operations (``unchange'', ``insert'', ``replace'', ``delete'') to each phone or word. For Word-Phone, 
the word is assigned to ``unchanged'' only when all the constituent phones are the same between phone-level alignments, otherwise the whole word is modified. 

An example of comparison among three correction methods is shown in Figure \ref{fig:correction_method_detail}. In this example, the target text is ``we are not happy'', but the actual pronounced text is ``we are heavily''. Moreover, the word ``we'' is pronounced to be ``W ER0'' instead of ``W IY''.
Different correction methods lead to different edit locations and operations. Phone-Phone and Word-Word methods determine edit locations and operations solely on the phone-level or word-level information respectively. In comparison, Word-Phone method determines edit locations and operations by incorporating both phone-level and word-level information. As word-level time stamps are usually more reliable than phone-level time stamps, performing correction on word-level is more desirable. 

Word-Word method is based on word-level alignment, while Phone-Phone and Word-Phone methods are based on phone-level alignment. Mispronunciation sometimes happens at phone-level rather than at word-level. Phone-Phone and Word-Phone methods are usually able to detect the mispronounced phones (``IY1'' is pronounced to be ``ER0'' in the example), but Word-Word method tends to neglect this type of phone-level mispronunciation error.

\subsection{Speech editing and correction (SE)}
With the edit locations and operations, the target text and the recorded speech are processed by a speech editing model to obtain the modified speech. We follow the network structure and editing process in EditSpeech \cite{tan2021editspeech}. We first predict and scale the duration of the target text. We retain the ``unchanged'' part of mel-spectrogram of the recorded speech.  We delete the speech segments marked as deletion from the original speech. Then we generate the modified mel-spectrogram for the frames that require replacement or insertion according to edit locations. Finally we convert the whole mel-spectrogram to speech waveform through a vocoder.

\vspace{-0.5em}
\section{Experimental setup} 
\label{sec:experiment}

\subsection{Datasets}
We perform experiments on two datasets, a manually perturbed version of VCTK \cite{yamagishi2019vctk} and L2-ARCTIC \cite{zhao2018l2arctic}. The perturbed VCTK simulates common mispronunciation in daily speaking. We apply the perturbed VCTK to evaluate speech correction performance of CorrectSpeech. We construct the perturbed VCTK by inserting, replacing and deleting word-level speech segments of the original VCTK speech utterances with 5\% probability for each operation. The speech segments for insertion and replacement are extracted from utterances of the same speaker. The S2T modules are less likely to utilize timbre change to detect perturbed location.  
L2-ARCTIC is a non-native English speech corpus. The speech utterances are spoken with strong accent. We apply the corpus to evaluate the accent reduction performance of CorrectSpeech.

\subsection{Model configuration}
We evaluate two S2T modules. For F-Align, we apply a pre-trained Conformer-CTC Large ASR model \cite{gulati2020conformer} with 121M parameters from NeMo \cite{kuchaiev2019nemo, nemo_conformer}
for word sequence recognition. 
This model is trained from thousands hours of English data. We then apply g2pE \cite{g2pE2019} to convert recognized word sequence to phone sequence. 
Finally, we apply GMM-HMM based Montreal Forced Aligner (MFA) \cite{mcauliffe2017montreal} to perform speech-text alignment and obtain word-level and phone-level time stamps. The frame rate of in MFA is 10 ms.
In CTC-Align, we train a phone-based Conformer-CTC Medium ASR model (27.3M parameters) with 100-hour ``train-clean-100'' subset of LibriSpeech \cite{panayotov2015librispeech} with NeMo. We retain stress symbols for phone labels (i.e., ``AH0'' and ``AH1'' are two different phones) in training. The frame rate is 40 ms due to sub-sampling in Conformer.

For T2T module, we apply the implementation of Needleman–Wunsch algorithm in the Biopython package \cite{biopython}
for phone-level and word-level alignment. For SE module, we apply EditSpeech \cite{tan2021editspeech} as our speech editing model and HiFi-GAN \cite{kong2020hifi} as our vocoder for waveform generation.

\section{Results and discussion} \label{sec:results}

\subsection{Objective evaluation}

\subsubsection{Phone recognition error}
We first analyze phone recognition errors of S2T modules. For perturbed VCTK dataset, the ground-truth perturbed phone sequences are the reference. For L2-ARCTIC, the manually annotated phone sequences are the reference. After removing the stress symbols, phone error rate (PER) is the performance metric. In addition, we include a ``stressed phone error rate" (s-PER) by retraining stress symbols (i.e., ``AH0'' and ``AH1'' are different). The results are shown in Table \ref{tab:recognition error}. The ASR model in F-Align achieves significantly lower recognition error than the ASR model in CTC-Align, probably due to more parameters and training data.

\begin{table}[h]
\centering
\caption{Phone recognition errors of S2T modules}
\label{tab:recognition error}
\scalebox{0.9}{
\begin{tabular}{c|c|c|c|c}
\hline
\multirow{2}{*}{metric} & \multicolumn{2}{c|}{perturbed VCTK} &
\multicolumn{2}{c}{L2-ARCTIC} \\
\cline{2-5}
~ & F-Align & CTC-Align & F-Align & CTC-Align\\
\hline
PER $\downarrow$ ($\%$) & 7.7 & 17.6 & 16.4 & 27.5\\
\hline
s-PER $\downarrow$ ($\%$) & 7.9 & 19.9 & 16.8 & 29.5\\
\hline
\end{tabular}
}
\vspace{-1em}
\end{table}

\subsubsection{Alignment error}
We analyze alignment errors made by the S2T modules. For the perturbed VCTK dataset, the ground-truth phone-level time stamps are the reference. For L2-ARCTIC, the manually annotated phone-level time stamps are reference. We follow the metrics in  \cite{sainath2020emitting} and calculate time gaps between automatically determined time stamps and the references. We also calculate the ratio of phones within tolerable time gaps ($\leq$ 100ms). The results are shown in Table \ref{tab:alignment error}. 

In F-Align, forced alignment is carried out with a GMM-HMM ASR model, which operates at higher frame rate. F-Align achieves average alignment error of 1 frame ($\leq 10 $ ms) for both corpora. For CTC-Align, the average alignment error for perturbed VCTK is also about 1 frames ($\leq 40 $ ms), which is still acceptable. However, there is a substantial performance degradation for non-native speech in L2-ARCTIC. The average alignment error is more than 1 frame ($> 40$ ms). More alignment errors occur beyond the tolerable range of 100 ms. The results suggest that for end-to-end CTC ASR model, higher frame rate is required to improve time resolution of alignment.

\begin{table}[h]
\centering
\caption{Alignment error of S2T modules}
\label{tab:alignment error}
\scalebox{0.85}{
\begin{tabular}{c|c|c|c|c}
\hline
\multirow{2}{*}{metric} & \multicolumn{2}{c|}{perturbed VCTK} &
\multicolumn{2}{c}{L2-ARCTIC} \\
\cline{2-5}
~ & F-Align & CTC-Align & F-Align & CTC-Align\\
\hline
\makecell[c]{average start\\gap$\downarrow$ (ms)} & 2.4 & 36.3 & 7.3 & 46.2\\
\hline
\makecell[c]{average end\\gap$\downarrow$ (ms)} & 2.3 & 37.0 & 7.7 & 50.9\\
\hline
\makecell[c]{tolerable start gap\\phone ratio$\uparrow$ ($\%$)} & 99.5 & 98.5 & 99.0 & 92.8\\
\hline
\makecell[c]{tolerable end gap\\phone ratio$\uparrow$ ($\%$)} & 99.5 & 97.3 & 99.0 & 91.2\\
\hline
\end{tabular}
}
\vspace{-1em}
\end{table}

\subsubsection{Editing error}
\label{sec:Editingerror}
We utilize mel-cepstral distortion (MCD) as the objective metric to evaluate performance of speech correction of different systems with the perturbed VCTK. We evaluate with different combinations of the S2T modules and the correction methods. In Oracle system, we use the ground-truth perturbed location and perturbed content for correction. The results are shown in Table \ref{tab:edit error}. Note that CTC-Align only recognizes phone sequence, where Word-Word correction method is not applicable.

\begin{table}[h]
\centering
\caption{MCD of different combinations of S2T modules and correction methods}
\label{tab:edit error}
\scalebox{0.92}{
\begin{tabular}{c|c|c|c|c}
\hline
\multicolumn{2}{c|}{\multirow{2}{*}{MCD$\downarrow$}} & \multicolumn{3}{c}{S2T modules
} \\
\cline{3-5}
\multicolumn{2}{c|}{} & Oracle & F-Align & CTC-Align\\
\hline
\multirow{3}{*}{\makecell[c]{correction\\methods}} & Word-Word & 4.58 & 4.80 & - \\
\cline{2-5}
~ & Word-Phone & 5.06 & 5.12 & 5.67 \\
\cline{2-5}
~ & Phone-Phone & 5.49 & 5.55 & 5.98 \\
\hline
\end{tabular}
}
\vspace{-0.5em}
\end{table}

The S2T modules play a significant role in spectral quality of corrected speech.
Systems based on F-Align are slightly inferior to the Oracle. The results suggest that a S2T module with low ASR error and highly accurate alignment guarantees good spectral quality of  corrected speech. Systems based on CTC-Align perform worse than those based on F-Align by a significant margin. As most of alignment errors of CTC-Align is still in acceptable range, the lower ASR accuracy plays a more significant role for performance degradation. We also observe that systems with Word-Word method perform the best. The performance of systems with Word-Phone method are in the middle. The performance of systems with the Phone-Phone method is inferior to other methods. 

\subsection{Subjective evaluation}

\subsubsection{Comparison of S2T modules}
\label{comparisonS2T}
We apply Word-Phone as the correction method on the perturbed VCTK. We compare different S2T modules and the Oracle system in terms of subjective quality of the corrected speech. 
In MOS tests, listeners are required to compare different systems on naturalness and content similarity between corrected speech and unperturbed original speech. Content similarity is defined as whether the uttered words are the same. In ABX test, listeners are required to show their preference  on the corrected speech utternaces of different systems. 18 listeners participate the test. MOS and ABX results are shown in Table \ref{tab:reconition-alignment modules MOS} and Table \ref{tab:reconition-alignment modules ABX} respectively.

\begin{table}[h]
\centering
\caption{MOS of S2T module with 95\% confidence interval (CI)}
\label{tab:reconition-alignment modules MOS}
\scalebox{0.96}{
\begin{tabular}{c|c|c|c}
\hline
MOS$\uparrow$ & Oracle & F-Align & CTC-Align\\
\hline
naturalness & 3.91$\pm$0.19 & 3.8$\pm$0.16 & 3.32$\pm$0.21\\
\hline
\makecell[l]{content similarity\\ to original speech} & 4.08$\pm$0.20 & 4.06$\pm$0.18 & 3.52$\pm$0.22\\
\hline
\end{tabular}
}
\vspace{-1em}
\end{table}

\begin{table}[h]
\centering
\vspace{-0.5em}
\caption{ABX test of S2T modules}
\label{tab:reconition-alignment modules ABX}
\begin{tabular}{c|c|c|c|c}
\hline
\multirow{2}{*}{system A} & \multirow{2}{*}{system B}& \multicolumn{3}{c}{preference ($\%$)}\\
\cline{3-5}
~ & ~ & A & B & no preference\\
\hline
Oracle & F-Align & 30.0 & 26.7 & 43.3\\
\hline
Oracle & CTC-Align & 62.2 & 23.3 & 14.5\\
\hline
F-Align & CTC-Align & 71.1 & 21.1 & 7.8\\
\hline
\end{tabular}
\vspace{-0.5em}
\end{table}

We observe that F-Align performs better than CTC-Align in terms of MOS. The results agree with the objective MCD results. The ABX test also suggests that the listeners prefer F-Align to CTC-Align. The performance gap between F-Align and Oracle is relatively small in terms of MOS. In the ABX test, 43.3\% of listeners show no preference between the two systems. The results suggest that F-Align is a reliable S2T module for speech correction.

\subsubsection{Comparison of correction methods} \label{sec:correction methods comparison}
We set the S2T module to F-Align. We compare the quality of corrected speech with the three different correction methods. We follow the testing protocol as described in Section \ref{comparisonS2T}. The same 18 listeners participate in the test. MOS and ABX results are shown in Table \ref{tab:correction methods MOS} and Table \ref{tab:correction methods ABX} respectively.

\begin{table}[h]
\centering
\caption{MOS of correction methods with 95\% CI}
\label{tab:correction methods MOS}
\scalebox{0.9}{
\begin{tabular}{c|c|c|c}
\hline
MOS$\uparrow$ & Word-Word & Word-Phone &Phone-Phone \\
\hline
naturalness & 3.96$\pm$0.19 & 3.73$\pm$0.22& 3.69$\pm$0.19 \\
\hline
\makecell[l]{content similarity\\ to original speech} & 4.12$\pm$0.20 & 3.89$\pm$0.23& 3.96$\pm$0.23 \\
\hline
\end{tabular}
}
\vspace{-0.5em}
\end{table}

\begin{table}[h]
\centering
\caption{ABX test of correction methods}
\label{tab:correction methods ABX}
\scalebox{0.95}{
\begin{tabular}{c|c|c|c|c}
\hline
\multirow{2}{*}{system A} & \multirow{2}{*}{system B}& \multicolumn{3}{c}{preference ($\%$)}\\
\cline{3-5}
~ & ~ & A & B & no preference\\
\hline
Word-Word & Word-Phone & 44.4 & 15.6 & 40.0\\
\hline
Word-Word & Phone-Phone & 60.0 & 18.9 & 21.1\\
\hline
Word-Phone & Phone-Phone & 45.6 & 31.1 & 23.3\\
\hline
\end{tabular}
}
\vspace{-1em}
\end{table}

The MOS of Word-Word method is the best on both naturalness and content similarity. 
For naturalness, Word-Phone achieves slightly better
MOS than Phone-Phone. For content similarity, Phone-Phone achieves better MOS than Word-Phone. From the ABX test, we observe mixed preference to the Word-Word and Word-Phone methods. While 44.4\% of listeners suggest that Word-Word is better, another 40.0\% of listeners think that there is no difference. The ABX test also suggests that the listeners prefer less to Phone-Phone method.

\subsection{Accent reduction application}
We evaluate CorrectSpeech with accent reduction application on L2-ARCTIC. Accent is likely due to inaccurate pronunciation at phone-level. We therefore evaluate on Word-Phone and Phone-Phone methods. We apply F-Align as the S2T module. For the MOS test, we follow the testing protocol as described in Section \ref{comparisonS2T}. In ABX test, listeners are required to show their preference between the original accented speech and the modified speech. The same 18 listeners participate in the test. The results are shown in Table \ref{tab:accent reduction system} and Table \ref{tab:accent reduction system ABX} for MOS and ABX respectively. 

\begin{table}[h]
\centering
\caption{MOS of accent reduction systems with 95\% CI}
\label{tab:accent reduction system}
\scalebox{0.95}{
\begin{tabular}{c|c|c}
\hline
 MOS$\uparrow$ &\makecell[c]{Word-Phone}  &\makecell[c]{Phone-Phone} \\ 
\hline
naturalness & 3.74$\pm$0.18 & 3.66$\pm$0.19 \\
\hline
\makecell[l]{content similarity \\to originial speech} & 3.78$\pm$0.19 & 3.89$\pm$0.20 \\
\hline
\end{tabular}
}
\vspace{-0.5em}
\end{table}

\begin{table}[h]
\vspace{-0.5em}
\centering
\caption{ABX test of accented speech and modified speech}
\label{tab:accent reduction system ABX}
\scalebox{1}{
\begin{tabular}{c|c|c|c}
\hline
\multirow{2}{*}{system }& \multicolumn{3}{c}{preference ($\%$)}\\
\cline{2-4}
~ &  accented & modified & no preference\\
\hline
Word-Phone  & 33.3 & 55.6  & 11.1\\
\hline
Phone-Phone  & 38.9 & 45.6 & 15.5\\
\hline
\end{tabular}
}
\vspace{-0.5em}
\end{table}

For naturalness, Word-Phone method again achieves better MOS than Phone-Phone method. For content similarity, Phone-Phone again achieves better MOS. The listeners tend to prefer the modified speech to original accented speech according to the ABX test results, with more preference to Word-Phone method. The results suggest that CorrectSpeech is promising for accent reduction scenario.

\vspace{-1em}
\section{Conclusion} 
\label{sec:conclusion}
We propose an automated mispronounced speech correction system named CorrectSpeech.
Taking target text and recorded speech as input, CorrectSpeech is able to carry out speech mispronunciation correction in a fully automated manner. Ablation study shows that the speech should be modified at word-level with either word-level or phone-level alignment for better naturalness and content similarity. Currently, the pipeline S2T module F-Align performs significantly better than the end-to-end CTC-Align module. Experimental results show that an end-to-end S2T model is still feasible, and a CTC-based ASR model with high phone accuracy and higher frame rate is recommended. Experiments on L2-ARCTIC demonstrate that CorrectSpeech is also applicable to speech accent reduction. We consider applying CorrectSpeech to more languages as our future work.


\vfill\pagebreak

\bibliographystyle{IEEEtran}

\bibliography{mybib}

\end{document}